\documentclass[conference]{IEEEtran}

\usepackage{graphicx}  

\begin{document}

\title{Testing Lorentz Invariance using Zeeman Transitions in Atomic Fountains}

\author{\authorblockN{Peter Wolf}
\authorblockA{SYRTE, Observatoire de Paris\\ and\\ Bureau International des Poids et Mesures\\
Pavillon de Breteuil, 92312 S\`evres, Cedex, France\\}
\and
\authorblockN{Fr\'ederic Chapelet\\
S\'ebastien Bize\\
Andr\'e Clairon\\
}
\authorblockA{SYRTE, Observatoire de Paris\\
61 av. de l'Observatoire, 75014 Paris, France\\
}}


%


\maketitle

\begin{abstract}
Lorentz Invariance (LI) is the founding postulate of Einstein's 1905 theory of relativity, and therefore at the heart of all accepted theories of physics. It characterizes the invariance of the laws of physics in inertial frames under changes of velocity or orientation. This central role, and indications from unification theories \cite{KostoSam,Damour1,Gambini} hinting toward a possible LI violation, have motivated tremendous experimental efforts to test LI. A comprehensive theoretical framework to describe violations of LI has been developed over the last decade \cite{Kosto1}: the Lorentz violating Standard Model Extension (SME). It allows a characterization of LI violations in all fields of present day physics using a large (but finite) set of parameters which are all zero when LI is satisfied. All classical tests (e.g. Michelson-Morley or Kennedy-Thorndike experiments \cite{Stanwix,Wolf2004}) can be analyzed in the SME, but it also allows the conception of new types of experiments, not thought of previously. We have carried out such a conceptually new LI test, by comparing particular atomic transitions (particular orientations of the involved nuclear spins) in the $^{133}$Cs atom using a cold atomic fountain clock. This allows us to test LI in a previously largely unexplored region of the SME parameter space, corresponding to first measurements of four proton parameters and an improvement by 11 and 12 orders of magnitude on the determination of four others. In spite of the attained accuracies, and of having extended the search into a new region of the SME, we still find no indication of LI violation.
\end{abstract}


%
\IEEEpeerreviewmaketitle

The experiment reported here tests for a violation of LI by searching for a variation of atomic transition frequencies as a function of the orientation of the spin of the involved atomic states (clock comparison or Hughes-Drever experiment). Based on the SME analysis of numerous atomic transitions in \cite{KL,Bluhm} we have chosen the measurement of a particular combination of transitions in the $^{133}$Cs atom. This provides good sensitivity to the quadrupolar SME energy shift of the proton (as defined in \cite{KL}) in the spin $\frac{7}{2}$ Cs nucleus, whilst being largely insensitive to magnetic perturbations (first order Zeeman effect). The corresponding region of the SME parameter space has been largely unexplored previously, with first limits on some parameters set only very recently \cite{Lane2005} by a re-analysis of the Doppler spectroscopy experiment (Ives-Stilwell experiment) of \cite{Saathoff}. Given the large improvement (11 and 12 orders of magnitude on four parameters) we obtain with respect to those results, and the qualitatively new region explored (first measurements of four parameters), the experiment had comparatively high probability for the detection of a positive Lorentz violating signal. However, our results clearly exclude that possibility.

The SME was developed relatively recently by Kosteleck\'y and co-workers \cite{Kosto1}, motivated initially by possible Lorentz violating phenomenological effects of string theory \cite{KostoSam}. It consists of a parameterized version of the standard model Lagrangian that includes all Lorentz violating terms that can be formed from known fields, and includes (in its most recent version) gravity. In its minimal form the SME characterizes Lorentz violation in the matter sector by 44 parameters per particle \cite{KL,Bluhm}, of which 40 are accessible to terrestrial experiments at first order in $v_\oplus/c$ \cite{Bluhm} ($v_\oplus$ is the orbital velocity of the Earth and $c = 299792458$ m/s). All SME parameters vanish when LI is verified. Existing bounds for the proton (p$^+$), neutron (n) and electron (e$^-$) come from clock comparison and magnetometer experiments using different atomic species (see \cite{KL} and references therein, \cite{Phillips,Bear,Hou,Cane}), from resonator experiments \cite{Muller2005,Wolf2004,Muller}, and from Ives-Stilwell experiments \cite{Lane2005,Saathoff}. They are summarized in tab. \ref{SME_tab}, together with the results reported in this work.

\begin{table*}
\caption{Orders of magnitude of present limits (in $\rm{GeV}$) on Lorentz violating parameters in the
minimal SME matter sector and corresponding references. Indeces $J, K$ run over $X,Y,Z$ with $J \neq K$. Limits from the present work are in bold type, with previous limits (when available) in brackets.}
\begin{center}
\begin{tabular}{ccccc}
\hline
Parameter & p$^+$ & n & e$^-$ & Ref.\\
\hline
$\tilde{b}_X$, $\tilde{b}_Y$ & $10^{-27}$ & $10^{-31}$ & $10^{-29}$ & \cite{Phillips}, \cite{Bear}, \cite{Hou} \\
$\tilde{b}_Z$ & - & - & $10^{-28}$ & \cite{Hou} \\
$\tilde{b}_T$, $\tilde{g}_T$, $\tilde{H}_{JT}$, $\tilde{d}_\pm$ & - &  $10^{-27}$ & - & \cite{Cane}\\
$\tilde{d}_Q$, $\tilde{d}_{XY}$, $\tilde{d}_{YZ}$ & - &  $10^{-27}$ & - & \cite{Cane}\\
$\tilde{d}_X$, $\tilde{d}_Y$ & $10^{-25}$ & $10^{-29}$ & $10^{-22}$ & \cite{KL}, \cite{Cane}, \cite{KL}\\
$\tilde{d}_{XZ}$, $\tilde{d}_{Z}$ & - & - & - \\
$\tilde{g}_{DX}$, $\tilde{g}_{DY}$ & $10^{-25}$ & $10^{-29}$ & $10^{-22}$ &\cite{KL}, \cite{Cane}, \cite{KL}\\
$\tilde{g}_{DZ}$, $\tilde{g}_{JK}$ & - & - & - \\
$\tilde{g}_{c}$ & - & $10^{-27}$ & - & \cite{Cane}\\
$\tilde{g}_{-}$, $\tilde{g}_{Q}$, $\tilde{g}_{TJ}$ & - & - & - \\
$\tilde{c}_{Q}$ & ${\bf 10}^{{\bf -22}(-11)}$ & - & $10^{-9}$ & \cite{Lane2005,Saathoff} \\
$\tilde{c}_X$, $\tilde{c}_Y$ & ${\bf 10^{-24}}$ & $10^{-25}$ & $10^{-19}$ &  \cite{KL}, \cite{Muller2005,Wolf2004,Muller} \\
$\tilde{c}_Z$, $\tilde{c}_{-}$ & ${\bf 10^{-24}}$ & $10^{-27}$ & $10^{-19}$ &  \cite{KL}, \cite{Muller2005,Wolf2004,Muller} \\
$\tilde{c}_{TJ}$ & ${\bf 10}^{{\bf -20}(-8)}$ & - & $10^{-6}$ & \cite{Lane2005,Saathoff} \\
\hline
\end{tabular}
\end{center}
\label{SME_tab}
\end{table*}

For our experiment we use one of the laser cooled fountain clocks operated at the Paris observatory, the $^{133}$Cs and $^{87}$Rb double fountain FO2 (see \cite{BizeJPB} for a detailed description). We run it in Cs mode on the $|F=3\rangle \leftrightarrow |F=4\rangle$ hyperfine transition of the $6S_{1/2}$ ground state. Both hyperfine states are split into Zeeman substates $m_F=[-3,3]$ and $m_F=[-4,4]$ respectively. The clock transition used in routine operation is $|F=3,m_F=0\rangle\leftrightarrow |F=4,m_F=0\rangle$ at 9.2 GHz, which is magnetic field independent to first order. The first order magnetic field dependent Zeeman transitions ($|3,i\rangle\leftrightarrow |4,i\rangle$ with $i=\pm 1,\pm 2,\pm 3$) are used regularly for measurement and characterization of the magnetic field, necessary to correct the second order Zeeman effect of the clock transition. In routine operation the clock transition frequency stability of FO2 is $1.6\times 10^{-14}\tau^{-1/2}$, and its accuracy $7\times 10^{-16}$ \cite{BizeJPB,Marion}.

A detailed description of the minimal SME as applied to the perturbation of atomic energy levels and transition frequencies can be found in \cite{KL,Bluhm}. Based on the Schmidt nuclear model\footnote{As discussed in \cite{KL} the Schmidt nuclear model only allows an approximate calculation of the SME frequency shift, with more complex models generally leading to dependences on additional SME parameters. Nonetheless, the model is sufficient to derive the leading order terms and has been used for the analysis of most experiments providing the bounds in Tab. \ref{SME_tab}.} one can derive the SME frequency shift of a Cs $|3,m_F\rangle\leftrightarrow |4,m_F\rangle$ transition in the form

\begin{eqnarray}
\label{clockshift}
\delta\nu &=& \frac{m_F}{14 h}\left(\beta_p\tilde{b}_3^p - \delta_p\tilde{d}_3^p + \kappa_p\tilde{g}_d^p\right)-\frac{m_F^2}{14 h}\left(\gamma_p\tilde{c}_q^p\right) \nonumber\\
&-& \frac{m_F}{2 h}\left(\beta_e\tilde{b}_3^e - \delta_e\tilde{d}_3^e + \kappa_e\tilde{g}_d^e\right)\\
&+& m_F K_Z^{(1)}B+\left(1-\frac{m_F^2}{16}\right)K_Z^{(2)}B^2\nonumber
\end{eqnarray}
for the quantization magnetic field in the negative z direction (vertically downward) in the lab frame. The first three terms in (\ref{clockshift}) are Lorentz violating SME frequency shifts, the last two describe the first and second order Zeeman frequency shift (we neglect $B^3$ and higher order terms). The tilde quantities are linear combinations of the SME matter sector parameters of tab. \ref{SME_tab} in the lab frame, with p,e standing for the proton and electron respectively. The quantities $\beta_w, \delta_w, \kappa_w, \gamma_w, \lambda_w$ (with $w=p,e$) depend on the nuclear and electronic structure, they are given in tab. II of \cite{Bluhm}, $h$ is Planck's constant, $B$ is the magnetic field seen by the atom, $K_Z^{(1)} = 7.008$~Hz~nT$^{-1}$ is the first order Zeeman coefficient, $K_Z^{(2)}=427.45\times 10^8$ Hz~T$^{-2}$ is the second order coefficient \cite{VanAud}. The tilde quantities in (\ref{clockshift}) are time varying due to the motion of the lab frame (and hence the quantization field) in a cosmological frame, inducing modulations of the frequency shift at sidereal and semi-sidereal frequencies, which can be searched for.

From (\ref{clockshift}) we note that the $m_F=0$ clock transition is insensitive to Lorentz violation or the first order Zeeman shift, while the Zeeman transitions ($m_F \neq 0$) are sensitive to both. Hence, a direct measurement of a Zeeman transition with respect to the clock transition allows a LI test. However, such an experiment would be severely limited by the strong dependence of the Zeeman transition frequency on $B$, and its diurnal and semi-diurnal variations. To avoid such a limitation, we "simultaneously" (see below) measure the $m_F=3$, $m_F=-3$ and $m_F=0$ transitions and form the combined observable $\nu_{c} \equiv \nu_{+3}+\nu_{-3}-2\nu_{0}$. From (\ref{clockshift})

\begin{equation}
\label{obsshift}
\nu_{c}= \frac{1}{7 h}K_p\tilde{c}_q^p-\frac{9}{8}K_{Z}^{(2)}B^2
\end{equation}
where we have used $\gamma_p = -K_p/9$ from \cite{Bluhm} ($K_p\sim 10^{-2}$ in the Schmidt nuclear model). This observable is insensitive to the first order Zeeman shift, and should be zero up to the second order Zeeman correction and a possible Lorentz violating shift in the first term of (\ref{obsshift}).  

The lab frame parameter $\tilde{c}_q^p$ can be related to the conventional sun-centered frame parameters of the SME (the parameters of tab. \ref{SME_tab}) by a time dependent boost and rotation (see \cite{Bluhm} for details). This leads to a general expression for the observable $\nu_c$ of the form

\begin{eqnarray}
\label{model}
\nu_{c}=A~&+&C_{\omega_\oplus}{\rm cos}(\omega_\oplus T_\oplus)+S_{\omega_\oplus}{\rm sin}(\omega_\oplus T_\oplus) \\
&+&C_{2\omega_\oplus}{\rm cos}(2\omega_\oplus T_\oplus)+S_{2\omega_\oplus}{\rm sin}(2\omega_\oplus T_\oplus), \nonumber
\end{eqnarray}
where $\omega_\oplus$ is the frequency of rotation of the Earth, $T_\oplus$ is time since 30 March 2005 11h 19min 25s UTC (consistent with the definitions in \cite{KM}), and with $A$, $C_{\omega_\oplus}$, $S_{\omega_\oplus}$, $C_{2\omega_\oplus}$ and $S_{2\omega_\oplus}$ given in tab. \ref{CS_tab} as functions of the sun frame SME parameters. A least squares fit of (\ref{model}) to our data provides the measured values given in tab. \ref{CS_tab}, and the corresponding determination of the SME parameters.

\begin{table*}
\caption{\label{CS_tab} Coefficients of (\ref{model}) to first order in $\beta \equiv v_\oplus/c$, where $\Omega_\oplus$ is the angular frequency of the Earth's orbital motion, $T$ is time since the March equinox, $\chi = 41.2^\circ$ is the colatitude of our lab, and $\eta=23.3^\circ$ is the inclination of the Earth's orbit. The measured values (in ${\rm mHz}$) are shown together with the statistical (first bracket) and systematic (second bracket) uncertainties. For our relatively short data set (21 days) we neglect the slow variation due to the annual terms and take $\Omega_\oplus T \sim 0.34$~${\rm rad.}$}
\begin{center}
\begin{tabular}{ccc}
\hline
$A$ & $\frac{K_p}{28h}\left(1+3{\rm cos}(2\chi)\right)\left(\tilde{c}_Q+\beta\left({\rm sin}(\Omega_\oplus T)\tilde{c}_{TX}+{\rm cos}(\Omega_\oplus T)\left(2{\rm sin}\eta~\tilde{c}_{TZ}-{\rm cos}\eta~\tilde{c}_{TY}\right)\right)\right)-\frac{9}{8}K_{Z}^{(2)}B^2$ & -5.7(0.05)(26) \vspace{1mm}\\
$C_{\omega_\oplus}$ & $-\frac{3K_p}{14h}{\rm sin}(2\chi)\left(\tilde{c}_Y+\beta\left({\rm sin}(\Omega_\oplus T)\tilde{c}_{TZ}-{\rm cos}(\Omega_\oplus T){\rm sin}\eta~\tilde{c}_{TX}\right)\right)$ & 0.1(0.07)(0.35) \vspace{1mm}\\
$S_{\omega_\oplus}$ & $-\frac{3K_p}{14h}{\rm sin}(2\chi)\left(\tilde{c}_X-\beta~{\rm cos}(\Omega_\oplus T)\left({\rm sin}\eta~\tilde{c}_{TY}+{\rm cos}\eta~\tilde{c}_{TZ}\right)\right)$ & -0.03(0.07)(0.35) \vspace{1mm}\\
$C_{2\omega_\oplus}$ & $-\frac{3K_p}{14h}{\rm sin}^2\chi\left(\tilde{c}_-+\beta\left({\rm sin}(\Omega_\oplus T)\tilde{c}_{TX}+{\rm cos}(\Omega_\oplus T){\rm cos}\eta~\tilde{c}_{TY}\right)\right)$ & 0.04(0.07)(0.35) \vspace{1mm}\\
$S_{2\omega_\oplus}$ & $-\frac{3K_p}{14h}{\rm sin}^2\chi\left(\tilde{c}_Z+\beta\left({\rm sin}(\Omega_\oplus T)\tilde{c}_{TY}-{\rm cos}(\Omega_\oplus T){\rm cos}\eta~\tilde{c}_{TX}\right)\right)$ & -0.02(0.07)(0.35) \\
\hline
\end{tabular}
\end{center}
\end{table*}

\begin{figure}[b]
\begin{center}
\includegraphics[width=7cm]{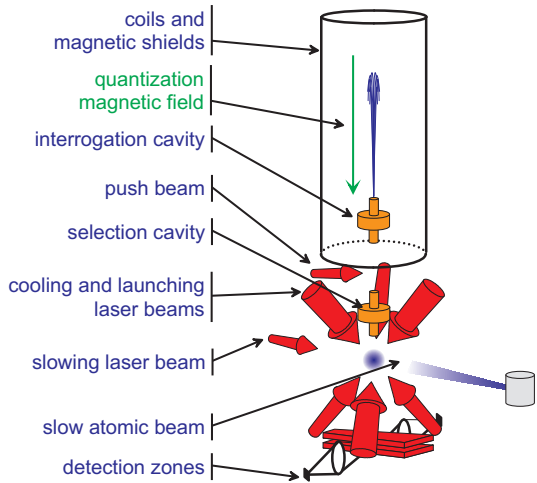}
\end{center}
\caption{Schematic view of an atomic fountain.} \label{fig:fountain}
\end{figure}

The FO2 setup is sketched in fig. \ref{fig:fountain}. Cs atoms
effusing from an oven are slowed using a chirped counter propagating laser
beam and captured in a lin $\perp$ lin optical molasses. Atoms are
cooled by six laser beams supplied by pre adjusted optical fiber couplers
precisely attached to the vacuum tank and aligned along the axes of
a 3 dimensional coordinate system, where the (111) direction is
vertical. Compared to typical FO2 clock operation \cite{BizeJPB}, the number of
atoms loaded in the optical molasses has been reduced to $2\times
10^{7}$ atoms captured in 30~ms. This reduces the collisional frequency shift of $\nu_c$ to below 0.1~mHz, and even less for its variation at $\omega_\oplus$ and $2\omega_\oplus$. 

Atoms are launched upwards at 3.94~m.s$^{-1}$ by using a moving
optical molasses and cooled to $\sim 1~\mu$K in the moving frame by
adiabatically decreasing the laser intensity and increasing the
laser detuning. Atoms are then selected by means of a microwave
excitation in the selection cavity performed in a bias magnetic
field of $\sim 20$~$\mu$T, and of a push laser beam. Any of the
$|3,m_F\rangle$ states can be prepared with a high
degree of purity (few $10^{-3}$) by tuning the selection microwave
frequency. 52~cm above the capture zone, a cylindrical copper cavity
(TE$_{011}$ mode) is used to probe the
$|3,m_F\rangle\leftrightarrow
|4,m_F\rangle$ hyperfine transition at 9.2~GHz. The
Ramsey interrogation method is performed by letting the atomic cloud
interact with the microwave field a first time on the way up and a
second time on the way down. After the interrogation, the
populations $N_{F=4}$ and $N_{F=3}$ of the two hyperfine levels are
measured by laser induced fluorescence, leading to a determination
of the transition probability $P=N_{3}/(N_{3}+N_{4})$ which is
insensitive to atom number fluctuations. One complete fountain cycle
from capture to detection lasts 1045~ms in the present experiment.
From the transition probability, measured on both sides of the
central Ramsey fringe, we compute an error signal to lock the
microwave interrogation frequency to the atomic transition using a
digital servo loop. The frequency corrections are applied to a
computer controlled high resolution DDS synthesizer in the microwave
generator. These corrections are used to measure the atomic
transition frequency with respect to the local reference signal used
to synthesize the microwave frequency.

The homogeneity and the stability of the magnetic field in the
interrogation region is a crucial point for the experiment. A
magnetic field of $203$~nT is produced by a main solenoid (length
815~mm, diameter 220~mm) and a set of 4 compensation coils.
These coils are surrounded by a first layer of 3 cylindrical
magnetic shields. A second layer is composed of 2 magnetic shields
surrounding the entire experiment (optical molasses and detection
zone included). Between the two layers, the magnetic field
fluctuations are sensed with a flux-gate magnetometer and stabilized
by acting on 4 hexagonal coils. The magnetic field in the
interrogation region is probed using the
$|3,1\rangle\leftrightarrow
|4,1\rangle$ atomic transition. Measurements of the transition frequency
as a function of the launch height show a peak to peak spatial
dependence of $230$~pT over a range of 320~mm above the interrogation
cavity with a variation of $\leq$ 0.1 pT/mm around the apogee of the atomic trajectories. Measurements of the same transition as a function of time at
the launch height of 791~mm show a magnetic field instability near
2~pT for an integration time of $\tau=$1~s. The long term behavior exhibits residual variations of the magnetic
field ($\sim$ 0.7~pT at $\tau=$10000~s) induced by temperature fluctuations which could cause
variations of the current flowing through solenoid, of the solenoid
geometry, of residual thermoelectric currents, of the magnetic
shield permeability, etc...

The experimental sequence is tailored to circumvent the limitation
that the long term magnetic field fluctuations could cause. First
$|3,-3\rangle$ atoms are selected and the
$|3,-3\rangle\leftrightarrow
|4,-3\rangle$ transition is probed at half maximum
on the red side of the resonance (0.528~Hz below the resonance
center). The next fountain cycle, $|3,+3\rangle$
atoms are selected and the
$|3,+3\rangle\leftrightarrow
|4,+3\rangle$ transition is also probed at half
maximum on the red side of the resonance. The third and fourth fountain cycles,
the same two transitions are probed on the blue side of the resonances (0.528~Hz above the resonance
centers). This 4180~ms long sequence is repeated so as to
implement two interleaved digital servo loops finding the line
centers of both the $|3,-3\rangle\leftrightarrow
|4,-3\rangle$ and the
$|3,+3\rangle\leftrightarrow
|4,+3\rangle$ transitions. Every 400 fountain cycles, the above sequence is interrupted and the
regular clock transition
$|3,0\rangle\leftrightarrow
|4,0\rangle$ is measured for 10~s allowing for an
absolute calibration of the local frequency reference with a
suitable statistical uncertainty. Using this sequence, magnetic field fluctuations over timescales $\geq 4$~s are rejected in the combined observable $\nu_c$ and the stability is dominated by the short term
($\tau < 4$~s) magnetic field fluctuations.

\begin{figure}[b]
\begin{center}
\includegraphics[width=7cm]{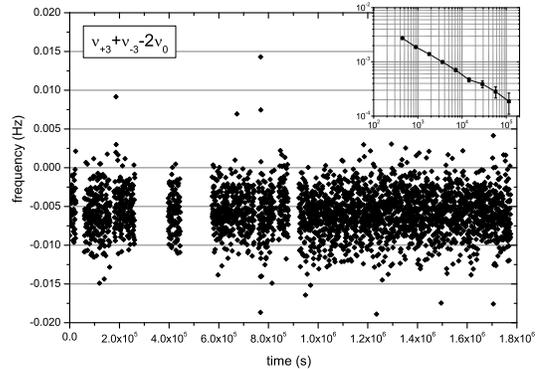}
\end{center}
\caption{Raw data of the measurements of $(\nu_{+3}+\nu_{-3}-2\nu_{0})$ spanning $\sim 21$ days. The inset shows the stability (Alan deviation) of the last continuous stretch of data ($\sim 10$ days).} \label{SMEclockPRL1}
\end{figure}

We have taken data implementing the experimental sequence described above over a period of 21 days starting on march 30, 2005. The complete raw data (no post-treatment) is shown in fig. \ref{SMEclockPRL1}, each point representing a $\sim$432~s measurement sequence of $\nu_{+3}+\nu_{-3}-2\nu_{0}$ as described above. The inset in fig. \ref{SMEclockPRL1} shows the frequency stability of the last continuous stretch of data ($\sim$10 days). We note the essentially white noise behavior of the data on that figure, indicating that the experimental sequence successfully rejects all long term variations of the magnetic field or of other perturbing effects. A least squares fit of the model (\ref{model}) to the complete data provides the 5 coefficients and statistical uncertainties given in tab. \ref{CS_tab}.

We note a statistically significant offset of the data from zero ($-5.7(0.05)$~mHz). This is partly due to the second order Zeeman shift (second term in (\ref{obsshift})) which amounts to $-2.0$~mHz. The remaining $-3.7$~mHz are either due to a systematic effect or indicate a genuine Lorentz violating signal.

The dominant systematic effect in our experiment is most likely a residual first order Zeeman shift. This arises when the $m_F=-3$ and $m_F=+3$ atoms have slightly different trajectories in the presence of a magnetic field gradient. The result is a difference in first order Zeeman shift and hence incomplete cancelation in the combined observable $\nu_c$. To test this hypothesis we have taken 3 hours of time of flight (TOF) measurements of the atoms in the different Zeeman states. We observe a significant TOF difference of $155~\mu$s (when launching at 4.25 m/s) between the $m_F=-3$ and $m_F=+3$ states, most likely due to slight magnetic and optical differences in the laser trapping and cooling during the final adiabatic phase. 

We model a residual first order Zeeman effect using a Monte Carlo (MC) simulation with the measured vertical magnetic field map, and assuming that the complete TOF difference is due to an initial vertical spatial offset ($\sim 660~\mu$m) between the atoms rather than an offset in launch velocity. In this scenario, which maximizes the resulting frequency shift, we obtain a frequency offset of only $\sim 5~\mu$Hz in $\nu_c$.

Another indication for magnetic field inhomogeneities comes from measurements of the contrast of the Ramsey fringes for the different Zeeman transitions. The measured contrasts are 0.94, 0.93, 0.87, and 0.75 for the $m_F=0$, $m_F=+1$, $m_F=+2$, and $m_F=+3$ transitions respectively. The observed differences cannot be explained by the known vertical magnetic field gradient alone, but are due to an additional horizontal field gradient, probably caused by residual magnetization inside the magnetic shields or lack of symmetry in the shielding or compensation. Our MC simulation reproduces the measured contrasts when we include a constant horizontal gradient of $\sim 6$~pT/mm. To check this value we have slightly tilted the fountain (by up to $\sim 1800~\mu$rad off the vertical) and measured the corresponding change in frequency of the $m_F=+1$ Zeeman transition. The obtained field gradient is somewhat smaller ($\sim 2$~pT/mm) but we consider the latter method less accurate, and conservatively use the larger value obtained from the contrast measurements.

Including the thus determined horizontal field gradient in a complete MC simulation we obtain a total offset in $\nu_c$ of $\sim 26$~mHz, assuming that the horizontal spatial offset between the $m_F=-3$ and $m_F=+3$ atoms is identical to the measured vertical offset ($660~\mu$m). We consider this to be an upper limit (the horizontal separation is likely to be less than the vertical one due to the absence of gravity), and take it as the systematic uncertainty on the determined offset ($A$ in tab. \ref{CS_tab}).

To determine the systematic uncertainty on the sidereal and semi-sidereal modulations of $\nu_c$ we have measured the variation of the $m_F=0$ TOF at those frequencies and taken the result as the maximum variation of the $m_F=+3$ vs $m_F=-3$ TOF difference (note that the real value is probably smaller, due to common mode variations). The best fit amplitudes of the sidereal and semi-sidereal TOF variation are 3.0~$\mu$s and 2.9~$\mu$s respectively, which leads to an upper limit of 0.35~mHz on the systematic uncertainties of $C_{\omega_\oplus}$, $S_{\omega_\oplus}$, $C_{2\omega_\oplus}$, $S_{2\omega_\oplus}$ in tab. \ref{CS_tab}.

Finally we use the five measurements and the relations in tab. \ref{CS_tab} to determine the values of the eight SME parameters. In doing so, we assume that there is no correlation between the three $\tilde{c}_{TJ}$ parameters and the other five parameters, and determine them independently. The results are given in tab. \ref{results}.

\begin{table}
\caption{\label{results}Results for SME Lorentz violating parameters $\tilde{c}$ for the proton, in ${\rm GeV}$ and with $J=X,Y,Z$.}
\begin{center}
\begin{tabular}{l}
\hline
$\tilde{c}_{Q\ }=-0.3(2.2)$ $\times 10^{-22}$ \hspace{5mm} $\tilde{c}_{-}=-0.2(1.6)$ $\times 10^{-24}$ \\
$\tilde{c}_{J\ }=\left(0.06(0.7),\ -0.2(0.7),\ \ 0.1(1.6)\right)$ $\ \times 10^{-24}$  \\
$\tilde{c}_{TJ}=\left(0.1(1.3),\ -0.04(1.7),\ -0.1(1.0)\right)$ $\times 10^{-20}$
\\
\hline
\end{tabular}
\end{center}
\end{table}

In conclusion, we have carried out a test of Lorentz invariance in the matter sector of the minimal SME using Zeeman transition in a cold $^{133}$Cs atomic fountain clock. We see no indication of a violation of LI at the present level of experimental uncertainty. Using a particular combination of the different atomic transitions we have set first limits on four proton SME parameters and improved previous limits \cite{Lane2005} on four others by 11 and 12 orders of magnitude.

Continuing the experiment regularly over a year or more will allow statistical decorrelation of the three $\tilde{c}_{TJ}$ parameters from the other five, due to their modulation at the annual frequency ($\Omega_\oplus T$ terms in tab. \ref{CS_tab}). Further improvements, and new measurements, could come from the unique capability of our fountain clock to run on $^{87}$Rb and $^{133}$Cs simultaneously. The different sensitivity of the two atomic species to Lorentz violation (see \cite{Bluhm}) and magnetic fields, should allow a measurement of all SME parameters in (\ref{clockshift}) in spite of the presence of the first order Zeeman effect. Ultimately, space clocks, like the planned ACES mission \cite{ACES} will provide the possibility of carrying out similar experiments but with faster (90 min orbital period) modulation of the putative Lorentz violating signal, and correspondingly faster data integration.

\end{document}